\documentclass[11pt,english,a4paper]{article}

\usepackage{amsmath, amssymb, amsthm, mathrsfs}
\usepackage[dvips]{geometry}
\usepackage[retainorgcmds]{IEEEtrantools}

\title{Monotonicity Results for Coherent MIMO Rician Channels}

\author{Daniel H\"osli \and Young-Han Kim \and Amos Lapidoth%
  \thanks{Daniel H\"osli and Amos Lapidoth are with the Information
    and Signal Processing Laboratory, ETH~Zurich, Switzerland.
    Young-Han Kim is with the Information Systems Laboratory, Stanford
    University, Stanford, CA 94305-9510, USA. The material in this
    paper was presented in part at the IEEE International Symposium on
    Information Theory, Yokohama, Japan, June/July 2003, and at the
    International ITG Conference on Source and Channel Coding,
    Erlangen, Germany, January 2004.}}

\newtheorem{definition}{Definition}[section]
\newtheorem{theorem}[definition]{Theorem}
\newtheorem{lemma}[definition]{Lemma}
\newtheorem{corollary}[definition]{Corollary}
\newtheorem{proposition}[definition]{Proposition}

\newcommand{\vect}[1]{\boldsymbol{\mathbf{#1}}}
\newcommand{\mat}[1]{\mathsf{#1}}
\newcommand{\vmat}[1]{\mathbb{#1}}                          
\newcommand{\trans}[1]{#1^{{\textnormal{\textsf{\tiny T}}}}}
\newcommand{\hermi}[1]{{#1}^{\dagger}}
\newcommand{\Unitary}[1]{\set{U}(#1)}
\newcommand{\Positive}[1]{\set{H}^+(#1)}
\newcommand{\I}{\mat{I}}

\newcommand{\trace}[1]{\operatorname{tr}\left(#1\right)} 
\newcommand{\diag}[1]{\operatorname{diag}\left(#1\right)}

\newcommand{\geql}{\succeq}
\newcommand{\leql}{\preceq}

\newcommand{\Complex}{\mathbb C}

\renewcommand{\d}{\,\textnormal{d}}
\newcommand{\Prv}[1]{\,\textnormal{Pr}\!\left[#1\right]}
\newcommand{\E}[2][]{\textnormal{\textsf{E}}_{#1}\!\left[#2\right]}
\newcommand{\set}[1]{\mathcal{#1}}
\newcommand{\NormalC}[2]{\mathcal{N}_{\Complex}\!\left({#1},{#2}\right)}
\newcommand{\eqlaw}{\stackrel{\mathscr{\scriptscriptstyle L}}{=}} % Equivalence in prob. law

\newcommand{\x}{\vect{x}}                 % realization
\newcommand{\X}{\vect{X}}                 % random input
\newcommand{\KX}{\mat{K}}                 % covariance matrix
\newcommand{\Es}{\set{E}}                 % input power
\newcommand{\Y}{\vect{Y}}                 % random channel output

\newcommand{\Z}{\vect{Z}}

\newcommand{\Hmat}{\vmat{H}}              % random fading matrix
\newcommand{\hmat}{\mat{H}}               % realization of the fading matrix
\newcommand{\D}{\mat{D}}                  % deterministic LoS matrix

\newcommand{\IF}{\mathcal{I}}     % mutual info induced by Gaussian input
\newcommand{\IIG}{I^{\rm{IG}}}    % mutual info induced by Gaussian input

\newcommand{\R}{R}

\begin{document}

\maketitle

\begin{abstract}
  The dependence of the Gaussian input information rate on the
  line-of-sight (LOS) matrix in multiple-input multiple-output
  coherent Rician fading channels is explored. It is proved that the
  outage probability and the mutual information induced by a
  multivariate circularly symmetric Gaussian input with any covariance
  matrix are monotonic in the LOS matrix $\mat{D}$, or more precisely,
  monotonic in $\hermi\D \D$ in the sense of the Loewner partial
  order.  Conversely, it is also demonstrated that this ordering on
  the LOS matrices is a necessary condition for the uniform
  monotonicity over all input covariance matrices.  This result is
  subsequently applied to prove the monotonicity of the isotropic
  Gaussian input information rate and channel capacity in the singular
  values of the LOS matrix. Extensions to multiple-access channels are
  also discussed.
\end{abstract}

\section{Introduction and Main Result}
\label{sec:introduction}
It is well known that the capacity of a single-input single-output
coherent Rician fading channel is monotonic in the magnitude of the
line-of-sight (LOS) component. This can be easily deduced from the
facts that the channel capacity is achieved by a zero-mean
circularly-symmetric Gaussian input and that a non-central chi-square
random variable is stochastically monotonic in the non-centrality
parameter \cite[Lemma~6.2~(b)]{lapidothmoser03_3},
\cite{johnsonkotzbalakrishnan95_2}. This result extends easily to the
single-input multiple-output and, with a little more work, to
multiple-input single-output scenarios, from the similar stochastic
monotonicity for the non-central chi-square random variable of a
higher degree.

The extension to the MIMO case, which may look straightforward at
first, requires some extra care, however.  The first difficulty one
encounters is that in order to demonstrate the monotonicity, one has
to introduce an ordering on the LOS matrices and it is \emph{a priori}
unclear what the natural ordering is for the problem at hand.  The
second difficulty is that there is no closed-form expression for the
capacity-achieving input distribution.  It is straightforward to
demonstrate that the capacity is achieved by a circularly-symmetric
multivariate Gaussian input, but no closed-form expressions for the
eigenvalues of the optimal covariance matrix are known. Finally, as in
the single-input case, under a fixed input distribution, one LOS
matrix may give rise to a larger information rate for a given
realization than another LOS matrix, but it may actually perform worse
when averaged over all fading realizations.

In this paper we show that the natural ordering on the LOS matrices
$\D$ is given by the Loewner partial order on $\hermi{\D}\D$, and
through this ordering we extend the monotonicity results to the MIMO
Rician channels.  More specifically, we say that the $m \times n$ LOS
matrix $\mat{D}$ is ``larger than or equal to'' the $m \times n$ LOS
matrix $\tilde{\mat{D}}$, if $\hermi{\mat{D}} \mat{D}$ is greater than
or equal to $\hermi{\tilde{\mat{D}}} \tilde{\mat{D}}$ in the Loewner
sense, i.e., if $\hermi{\mat{D}} \mat{D} - \hermi{\tilde{\mat{D}}}
\tilde{\mat{D}}$ is a positive semidefinite $n \times n$
matrix.\footnote{We point out that the Loewner partial order on
  $\hermi{\mat{D}} \mat{D}$ induces a \emph{preorder} on the LOS
  matrices $\D$, for $\hermi{\mat{D}} \mat{D} \geql
  \hermi{\tilde{\mat{D}}} \tilde{\mat{D}}$ and
  $\hermi{\tilde{\mat{D}}} \tilde{\mat{D}} \geql \hermi{\mat{D}}
  \mat{D}$ implies $\hermi{\mat{D}} \mat{D} = \hermi{\tilde{\mat{D}}}
  \tilde{\mat{D}}$, but \emph{not\,} $\tilde{\D} = \D$.  It only
  implies $\tilde{\D} = \mat{U} \D$ for some unitary matrix
  $\mat{U}$.}  (Here $\hermi\D$ is the Hermitian conjugate of $\D$.)
Under this ordering on the LOS matrices, we shall show the
monotonicity of channel capacity, the monotonicity of the isotropic
Gaussian information rate, and the monotonicity of outage probability.

We shall also extend the discussion to the multiple-access channel
(MAC). The MAC poses an additional challenge in that the capacity
region depends not only on the LOS matrices of different users
individually, but also on how these matrices relate to each other.
This requires a joint preorder on LOS matrices, as will be made clear
in the next section.

It should be emphasized that our monotonicity results are proved when
the distribution of the granular component is held fixed.
Consequently, as we vary the LOS matrix the output power is not held
fixed.  See \cite{cottatelluccidebbah04, jayaweerapoor05,
  lebrunfaulknershafismith04,driessenfoschini99} for studies where the
output power is held fixed.

We state our main result, from which the monotonicity results will follow.
\begin{theorem}
  \label{th:main}
  Let $\Hmat$ be a random $m \times n$ matrix whose components are
  independent, each with a zero-mean unit-variance circularly
  symmetric complex Gaussian distribution. If two deterministic
  complex $m \times n$ matrices $\D$, $\tilde{\D}$ are such that
  \begin{equation*}
    \hermi{\D} \D \geql \hermi{\tilde{\D}} \tilde{\D}
  \end{equation*}
  then we have
  \begin{IEEEeqnarray*}{rCl}
    \Prv{ \log\det\left( \mat{I}_m +
        (\Hmat + \D) \KX \hermi{(\Hmat + \D) } \right) \leq t}
    &\leq& \Prv{ \log\det\left( \mat{I}_m +
        (\Hmat + \tilde{\D}) \KX \hermi{(\Hmat + \tilde{\D}) } \right)
      \leq t}
  \end{IEEEeqnarray*}
for any $t \ge 0$ and any positive semidefinite $n\times n$ matrix
$\mat{K}$.
\end{theorem}

In this theorem and throughout, the notation $\mat{A} \geql \mat{B}$
indicates that $\mat{A} - \mat{B}$ is positive semidefinite.  The
notation $\mat{I}_m$ denotes the $m$-dimensional identity matrix.  We
use $\Positive{n}$ to denote the set of all $n \times n$ positive
semidefinite Hermitian matrices and use $\Unitary{n}$ for the set of
all unitary $n \times n$ matrices.  For a complex matrix $\mat{A}$,
$\trans{\mat{A}}$ denotes its transpose while $\hermi{\mat{A}}$
denotes its Hermitian conjugate (i.e., elementwise complex conjugate
of $\trans{\mat{A}}$).  We extend the usual notion of diagonality to
non-square matrices by saying that any matrix $\mat{A}$ is
\emph{diagonal} if $\mat{A}_{ij} = 0$ for all $i \ne j$.  All vectors
are column vectors unless specified otherwise.  All logarithms are
natural, i.e., to the base $e$.

In the following section we shall describe the single-user and the
multiple-access Rician fading channels and present the main corollaries
of Theorem~\ref{th:main}. The proof of Theorem~\ref{th:main} is given
in Section~\ref{sec:proof-monotonicity}.

%%%%%%%%%%%%%%%%%%%%%%%%%%%%%%%%%%%%%%%%%%%%%%%%%%%%%%%%%%%%%%%%%%%%%

\section{Applications}

We introduce two functions that will simplify the notation in our
subsequent discussion.  In the notation of Theorem~\ref{th:main}, we
define for any $t \geq 0$ and $\KX \in \Positive{n}$
\begin{equation*}
  F(t; \KX, \D) \triangleq \Prv{ \log\det\left( \mat{I}_m +
      (\Hmat + \D) \KX
      \hermi{(\Hmat + \D) } \right) \leq t }
\end{equation*}
and
\begin{equation*}
  \IF(\KX, \D) \triangleq 
  \E{\log\det\left( \I_m +
      (\Hmat + \D) \KX \hermi{(\Hmat + \D)}
    \right)}.
\end{equation*}
Noting that
\begin{equation}
  \IF(\KX, \D) = \int_{0}^{\infty} \big(1 - F(t; \KX, \D) \big) \d t
  \label{eq:I_to_F}
\end{equation}
we obtain the following corollary of Theorem~\ref{th:main}.
\begin{corollary}
  \label{coro:I_F}
  If $\hermi{{\D}} {\D} \geql \hermi{\tilde{\D}} \tilde{\D}$, then
  \begin{equation*}
    \IF(\KX, \D) \ge \IF(\KX, \tilde{\D}), \quad \forall \KX \in \Positive{n}.
  \end{equation*}
\end{corollary}

The following converse to Corollary~\ref{coro:I_F} also holds, which
shows that the preorder on the LOS matrices is natural:
\begin{proposition}
  \label{prop:converse}
  If $\IF(\KX, \D) \geq \IF(\KX, \tilde{\D})$ for all $\KX \in
\Positive{n}$, then $\hermi{{\D}} {\D} \geql \hermi{\tilde{\D}}
\tilde{\D}$.
\end{proposition}
\begin{proof}
  See Appendix~\ref{sec:proof-prop-refpr}.
\end{proof}

We further note the rotational symmetry in $F(t;\KX,\D)$ and
$\IF(\KX,\D)$.  First observe that the law of $\Hmat$ is invariant
under left and right rotations, i.e., for any $\mat{U}\in\Unitary{m}$
and $\mat{V}\in\Unitary{n}$,
\begin{equation*}
  \mat{U} \Hmat \hermi{\mat{V}} \eqlaw \Hmat.
\end{equation*}
Consequently, we have for any $\mat{U} \in \Unitary{m}$ and $\mat{V}
\in\Unitary{n}$
\begin{IEEEeqnarray}{rCl}
  \nonumber
  F(t ; \KX, \mat{U} \D \hermi{\mat{V}} ) 
  &=& \Prv{ \log\det\left( \mat{I}_m +
      (\Hmat + \mat{U}\D \hermi{\mat{V}}) \KX
      \hermi{(\Hmat + \mat{U} \D \hermi{\mat{V}})} \right) \leq t} \\
  \nonumber
  &=& \Prv{ \log\det\left( \mat{I}_m + (\mat{U}
      \Hmat \hermi{\mat{V}} + \mat{U}\D \hermi{\mat{V}}) \KX
      \hermi{(\mat{U} \Hmat \hermi{\mat{V}} + \mat{U}\D \hermi{\mat{V}})}
    \right) \leq t} \\
  \nonumber
  &=& \Prv{ \log\det\left( \mat{U} \left(\I_m + 
        (\Hmat + \D) \hermi{\mat{V}} \KX \mat{V} \hermi{( \Hmat + \D )}
      \right) \hermi{\mat{U}} \right) \leq t} \\
  \nonumber
  &=& \Prv{ \log\det\left( \I_m + 
      (\Hmat + \D) \hermi{\mat{V}} \KX \mat{V} \hermi{( \Hmat + \D )}
    \right)  \leq t} \\
  &=& F(t ; \hermi{\mat{V}} \KX \mat{V}, \D).
  \label{eq:rotF}
\end{IEEEeqnarray}
 From this and \eqref{eq:I_to_F}, we thus have
\begin{equation}
  \label{eq:rotI}
  \IF(\KX, \mat{U} \D \hermi{\mat{V}} ) =
  \IF(\hermi{\mat{V}} \KX \mat{V}, \D).
\end{equation}

\subsection{The Single-User Rician Fading Channel \label{sec:su}}

The output $(\Hmat, \vect{Y})$ of the coherent single-user Rician (or
Ricean in certain dialects) fading channel consists of a random $m
\times n$ matrix $\Hmat$ whose components are independent and
identically distributed (IID) according to the zero-mean unit-variance
circularly symmetric complex Gaussian distribution $\NormalC{0}{1}$,
and of a random $m$-vector $\Y \in \Complex^{m}$ given by
\begin{equation}
  \Y = (\Hmat + \D) \x + \Z
 \label{eq:channel_model}
\end{equation}
where $\x \in \Complex^n$ is the channel input; $\D$ is a
deterministic $m \times n$ complex 
LOS matrix; and $\Z \in \Complex^m$ is
drawn according to the zero-mean circularly symmetric complex
multivariate Gaussian distribution $\NormalC{\vect{0}}{\sigma^2\I_m}$
for some $\sigma^{2} > 0$.  It is assumed that $\Hmat$ and $\Z$ are
independent of each other, and that their joint law does not depend on
the channel input~$\x$.

Since the law of $\Hmat$ does not depend on $\x$, we can express the
mutual information between the channel input and output as
\begin{equation}
  \label{eq:cond_MI}
  I \bigl( \X; \Hmat, \Y \bigr) = I \bigl( \X ; \Y \big| \Hmat
  \bigr).
\end{equation}
Of all input distributions of a given covariance matrix, the zero-mean
circularly symmetric multivariate complex Gaussian maximizes the
conditional mutual information $I(\X ; \Y | \Hmat = \hmat )$,
irrespective of the realization $\Hmat = \hmat$. Consequently, it also
maximizes the average mutual information $I( \X ; \Y | \Hmat )$. We
shall therefore consider in this paper zero-mean circularly symmetric
Gaussian input distributions $\NormalC{\vect{0}}{\KX}$ only.
focus on the dependence of mutual information on the LOS matrix~$\D$
when the input covariance matrix $\KX$ is held fixed.  Also, since we
can absorb the dependence on $\sigma^2$ into $\KX$, we assume
$\sigma^2 = 1$ without loss of generality.

For a given realization $\Hmat = \hmat$, we can express the
conditional mutual information $I( \X ; \Y | \Hmat = \hmat )$ for a
$\NormalC{\vect{0}}{\KX}$ input as
\begin{equation}
  \label{eq:Gauss_cond_MI}
  I( \X ; \Y | \Hmat = \hmat ) = \log\det\left( \I_m +
    (\hmat + \D) \KX \hermi{(\hmat + \D)}
  \right).
\end{equation}
By taking the expectation with respect to $\Hmat$, we can express the
average conditional mutual information as an explicit function of
$\KX$ and $\D$ as
\begin{IEEEeqnarray*}{rCl}
  I( \X ; \Y | \Hmat )
  &=& \E{\log\det\left( \I_m +
      (\Hmat + \D) \KX \hermi{(\Hmat + \D)} \right)} \\
  &=& \IF(\KX, \D).
\end{IEEEeqnarray*}
Thus Corollary~\ref{coro:I_F} can be interpreted as the monotonicity
of the average conditional mutual information of the Rician fading
channel~\eqref{eq:channel_model} with fixed input covariance matrix.
We can also give a more direct interpretation of Theorem~\ref{th:main}
through the notion of \emph{outage probability}. Consider the
probability
\begin{equation*}
  \Prv{ \log\det\left( \I_m + (\Hmat + \D) \KX \hermi{( \Hmat + \D )}
    \right)  \leq R} = F(R; \KX, \D).
\end{equation*}
We can interpret this quantity as the probability that the realization
$\hmat$ of $\Hmat$ will be such that the information rate on the
Gaussian channel $\vect{Y} = (\D + \hmat) \x + \vect{Z}$ for the input
distribution $\NormalC{0}{\KX}$ does not exceed $\R$. Under this
interpretation, Theorem~\ref{th:main} can be viewed as the
monotonicity of the outage probability in the channel LOS matrix.

These monotonicity results can be used to study the power-$\Es$
isotropic Gaussian input information rate
\begin{equation*}
  \IIG(\Es, \D) \triangleq \IF\left( \frac{\Es}{n}\I_n, \D \right)
\end{equation*}
and the capacity $C(\Es, \D)$ of the Rician channel under the average
input power constraint $\E{ \hermi{\X}\X } \le \Es$:
\begin{equation}
  \label{eq:capacity}
  C(\Es,\D) \triangleq \max_{\KX}\; \IF(\KX, \D)
\end{equation}
where the maximum is taken over the set of all input covariance
matrices~$\KX$ satisfying the trace constraint
\begin{equation}
  \label{eq:const}
  \trace{\KX} \leq \Es.
\end{equation}
It follows immediately from Corollary~\ref{coro:I_F} that, if
$\hermi{{\D}} {\D} \geql \hermi{\tilde{\D}} \tilde{\D}$, then
$\IIG(\Es, {\D})\geq \IIG(\Es, \tilde{\D})$ and $C(\Es, {\D})\geq
C(\Es, \tilde{\D})$.

Theorem~\ref{th:main} can also be used to study the rate-$\R$ outage
probability corresponding to the isotropic Gaussian input of power-$\Es$
\begin{equation*}
  P_{\textnormal{out}}^{\rm{IG}}(\R, \Es, \D) \triangleq
  F\left( \R, \frac{\Es}{n} \I_n, \D \right)
\end{equation*}
and the optimal power-$\Es$ rate-$\R$ outage probability
$P_{\textnormal{out}}^*(\R, \Es, \D)$, which is the smallest outage
probability that can be achieved for the rate $\R$ and the average
power $\Es$:
\begin{equation}
  \label{eq:defPout_star}
  P_{\textnormal{out}}^*(\R, \Es, \D) \triangleq \min_{\KX} F(\R, \KX, \D)
\end{equation}
where the minimum is over all positive semidefinite matrices~$\KX$
satisfying \eqref{eq:const}. From Theorem~\ref{th:main} we now obtain
that $\hermi{{\D}} {\D} \geql \hermi{\tilde{\D}} \tilde{\D}$ implies
that $P_{\textnormal{out}}^{\rm{IG}}(\R, \Es, \D) \le
P_{\textnormal{out}}^{\rm{IG}}(\R, \Es, \tilde{\D})$ and
$P_{\textnormal{out}}^*(\R, \Es, \D) \le P_{\textnormal{out}}^*(\R,
\Es, \tilde{\D})$.\footnote{Note that from the definition of power-$\Es$
  \emph{$\epsilon$-outage capacity} $C_{\textnormal{out}}^*(\epsilon,
  \Es, \D) \triangleq \sup \{ \R: P_{\textnormal{out}}^*(\R, \Es, \D)
  < \epsilon \}$ we immediately get the monotonicity
  $C_{\textnormal{out}}^*(\epsilon, \Es, \D) \geq
  C_{\textnormal{out}}^*(\epsilon, \Es, \tilde{\D})$ if $\hermi{{\D}}
  {\D} \geql \hermi{\tilde{\D}} \tilde{\D}$. A similar monotonicity
  holds for $C_{\textnormal{out}}^{\rm{IG}}(\epsilon, \Es, \D)
  \triangleq \sup \{ \R: P_{\textnormal{out}}^{\rm{IG}}(\R, \Es, \D) <
  \epsilon \}$.}

Using the rotational invariance~\eqref{eq:rotI}, we can strengthen
these results by stating them in terms of the singular values of the
LOS matrices. Indeed, for any unitary matrix $\mat{V}$, we have
$\trace{\hermi{\mat{V}} \KX \mat{V}} = \trace{\KX},$ and hence it
follows from \eqref{eq:rotI} that for any $\mat{U} \in \Unitary{m}$
and $\mat{V} \in \Unitary{n}$
\begin{equation*}
  \IIG(\Es, \mat{U} \D \hermi{\mat{V}}) = \IIG(\Es, \D)
\end{equation*}
and
\begin{equation*}
  C(\Es, \mat{U} \D \hermi{\mat{V}}) = C(\Es, \D)
\end{equation*}
i.e., that the isotropic Gaussian input information rate and channel
capacity depend on the LOS matrix only via its singular values.  By a
similar argument, it can be verified that, by \eqref{eq:rotF}, both
the outage probability corresponding to the isotropic Gaussian input
$P_{\textnormal{out}}^{\rm{IG}}(\R, \Es, \D)$ and the optimal outage
probability $P_{\textnormal{out}}^*(\R, \Es, \D)$ depend on the LOS
matrix $\D$ only via its singular values. Consequently, all these
quantities are monotonic in the singular values of the LOS matrix:
\begin{corollary}
  \label{cor:mono-cap}
  Let $\sigma_{1} \geq \sigma_{2} \geq \cdots \geq
  \sigma_{\min\{m,n\}}$ and $\tilde{\sigma}_{1} \geq
  \tilde{\sigma}_{2} \geq \cdots \geq \tilde{\sigma}_{\min\{m,n\}}$ be
  the singular values of the LOS matrices $\D$ and $\tilde{\D}$,
  respectively.  Suppose that $\sigma_i \ge \tilde{\sigma}_i$ for all
  $i$. Then
  \begin{IEEEeqnarray*}{rCl}
    \IIG(\Es, \D) &\geq& \IIG(\Es, \tilde{\D}) \\
    C(\Es, \D) &\geq& C(\Es, \tilde{\D}) \\
    P_{\textnormal{out}}^{\rm{IG}}(\R, \Es, \D) &\leq&
    P_{\textnormal{out}}^{\rm{IG}}(\R, \Es, \tilde{\D}) \\
    \noalign{\noindent and \vspace{\jot}}
    P_{\textnormal{out}}^*(\R, \Es, \D) &\leq&
    P_{\textnormal{out}}^*(\R, \Es, \tilde{\D}).
  \end{IEEEeqnarray*}
\end{corollary}

We can obtain an alternative proof (cf.~\cite{hoeslilapidoth04}) of
this corollary based on the observation that, if the LOS matrix $\D$
is diagonal, the capacity-achieving covariance matrix $\KX$ is also
diagonal.  (See also \cite{venkatesansimonvalenzuela03}.)  Since this
structural theorem on the capacity-achieving input distribution is of
independent interest, we restate it here.
\begin{theorem}
  \label{th:opt-covariance-structure}
  Suppose that $\hermi{\D}\D$ has the eigenvalue decomposition
  $\hermi{\D}\D = \mat{V} \mat{L} \hermi{\mat{V}}$ for some unitary
  matrix $\mat{V}$ and diagonal matrix $\mat{L}$.  Then the
  capacity-achieving covariance matrix~$\KX_*$ is given by
  \begin{equation*}
    \KX_* = \mat{V} \mat{\Lambda} \hermi{\mat{V}}
  \end{equation*}
  for some diagonal matrix $\mat{\Lambda}$.
\end{theorem}

\begin{proof}
  We show that if $\D$ is diagonal, the capacity-achieving input
  covariance matrix~$\KX_*$ is diagonal. The general case follows from
  \eqref{eq:rotI} and \eqref{eq:capacity}.
  
  Fix some $1 \le j \le n$.  Let $\mat{V} \in \Unitary{n}$ be a
  diagonal matrix with all diagonal entries equal to $1$ except the
  $j$-th entry, which is $-1$.  Similarly, let ${\mat{U}} \in
  \Unitary{m}$ be diagonal with all diagonal entries equal to $1$
  except for the $j$-th entry being $-1$.  (In case $j > m$,
  ${\mat{U}} = \I_m.$) Since $\mat{D}$ is diagonal, we have
  \begin{equation}
    \label{eq:diagI}
    {{\mat{U}}} \D \hermi{\mat{V}} = \D.
  \end{equation}

  Let $\tilde{\KX} = \hermi{\mat{V}} \KX {\mat{V}}$.  From
  \eqref{eq:diagI} and the rotational invariance~\eqref{eq:rotI}, we
  have
  \begin{IEEEeqnarray}{rCl}
    \IF(\tilde{\KX},\D) &=& \IF(\hermi{\mat{V}} \KX {\mat{V}},\D)
    \nonumber \\
    &=& \IF(\KX,{{\mat{U}}} \D \hermi{\mat{V}}) 
    \nonumber \\
    &=& \IF(\KX,\D)
    \label{eq:mut_info_K1}.
  \end{IEEEeqnarray}
  Now consider the matrix $\hat{\KX} = \frac{1}{2} \left(\KX +
    \tilde{\KX}\right)$. We note that the entries of $\hat{\KX}$ are
  identical to those of $\KX$ except that its off-diagonal elements in
  the $j$-th row and in the $j$-th column are zero.  In particular,
  $\trace{\KX} = \trace{\hat{\KX}}$.  On the other hand, it follows
  from \eqref{eq:mut_info_K1}, the strict concavity of $\IF(\KX,\D)$
  in $\KX$, and Jensen's inequality that
  \begin{IEEEeqnarray*}{rCl}
    \IF(\hat{\KX},\D) &\geq& \frac{1}{2} \left(\IF(\KX,\D) +
      \IF(\tilde{\KX},\D) \right) \\
    &=& \IF(\KX,\D)
  \end{IEEEeqnarray*}
  with equality if, and only if, $\KX = \hat{\KX}$. Repeating this
  procedure for each $j = 1,\ldots, n-1$ shows that an optimal
  covariance matrix must be diagonal.
\end{proof}

\subsection{The Rician Multiple-Access Fading Channel}

The coherent MIMO Rician multiple-access channel (MAC) with $k$ senders
is modeled as follows. The channel output consists of $k$ independent
random matrices $\Hmat_{1}, \ldots, \Hmat_{k}$, where $\Hmat_{i}$ is a
random $m \times n_{i}$ matrix whose components are IID
$\NormalC{0}{1}$, and of a random vector $\Y \in \Complex^{m}$ of the
form
\begin{equation}
  \Y = \sum_{i=1}^{k} (\Hmat_{i} + \D_{i} ) \x_i + \Z
  \label{eq:MAC_channel_model}
\end{equation}
where $\x_i \in \Complex^{n_i}$ is the $i$-th transmitter's input
vector, $\D_i$ is a deterministic $m \times n_i$ complex matrix
corresponding to the LOS matrix of the $i$-th user, and $\Z \sim
\NormalC{\vect{0}}{\sigma^2 \I_m}$ corresponds to the additive noise
vector.  It is assumed that all fading matrices $\{\Hmat_i\}_{i=1}^k$
are independent of~$\Z$ and that the joint distribution of
$(\Hmat_1,\ldots,\Hmat_k,\Z)$ does not depend on the inputs~$\{ \x_i
\}_{i=1}^k$. Without loss of generality, we will assume $\sigma^2 =
1$.

As in the single-user scenario, it can be shown~\cite{wyner94,
shamaiwyner97} that Gaussian inputs achieve the capacity region of the
MIMO Rician MAC.  The rate region $\set{R}(\KX_1,\ldots,\KX_k;
\D_1,\ldots,\D_k)$ achieved by independent Gaussian
inputs~$\NormalC{\vect{0}}{\KX_i}$ over the MIMO Rician MAC with LOS
matrices~$\{\D_i\}_{i=1}^k$ is given as the set of all rate vectors
$(\R_1,\ldots,\R_k)$ satisfying
\begin{equation}
  \label{eq:MAC-rate}
  \sum_{i \in \set{S}} \R_i \leq \E{\log\det\left( \I_m +
      \sum_{i \in \set{S}} (\Hmat_i + \D_i)
      \KX_i \hermi{(\Hmat_i + \D_i)} \right)} 
\end{equation}
for all $\set{S} \subseteq \{1, \ldots, k\}.$ The capacity region of
the MIMO Rician MAC, denoted as an explicit function of the input
power constraints on the different users and of their corresponding
LOS matrices, can be written as
\begin{equation}
  \set{C}(\Es_1, \ldots, \Es_k; \D_1, \ldots, \D_k) = 
  \bigcup_{\{\KX_i\}_{i=1}^k}
  \set{R}(\KX_1,\ldots, \KX_k; \D_1,\ldots,\D_k)
  \label{eq:MAC-capacity}
\end{equation}
where the union is over all input covariance matrices
$\{\KX_i\}_{i=1}^k$ that satisfy the trace constraints $\trace{\KX_i}
\leq \Es_i$, $i=1, \ldots, k$.

For each set $\set{S} \subseteq \{1,\ldots,k\}$ of elements $1 \leq
i_1 < i_2 < \ldots < i_s \leq k$, define the block matrices
\begin{IEEEeqnarray*}{rCl}
  \D_{\set{S}} &\triangleq& [\D_{i_1}, \ldots, \D_{i_s}] \\
  \Hmat_{\set{S}} &\triangleq& [\Hmat_{i_1}, \ldots, \Hmat_{i_s}]\\
  \noalign{\noindent and \vspace{\jot}}
  \KX_{\set{S}} &\triangleq& \diag{\KX_{i_1} , \ldots, \KX_{i_s}}.
\end{IEEEeqnarray*}
Further define $\D = [\D_1, \ldots, \D_k]$. Under this simplified
notation, the rate region~\eqref{eq:MAC-rate} can be expressed as
\begin{equation*}
  \sum_{i \in \set{S}} \R_i \leq \E{\log\det\left( \I_m +
      (\Hmat_\set{S} + \D_\set{S})
      \KX_\set{S} \hermi{(\Hmat_\set{S} + \D_\set{S})} \right)} = 
  \IF(\KX_\set{S}, \D_\set{S}).
\end{equation*}
Since the condition $\hermi{\D}\D \geql \hermi{\tilde{\D}}\tilde{\D}$
implies that $\hermi{\D_\set{S}}\D_\set{S} \geql
\hermi{\tilde{\D}_\set{S}}\tilde{\D}_\set{S}$ for all $\set{S}
\subseteq \{1,\ldots,k\}$, it follows from Corollary~\ref{coro:I_F}
that
\begin{equation*}
  \set{R}(\KX_{1}, \ldots, \KX_{k}; {\D}_{1}, \ldots, {\D}_{k})
  \supseteq 
  \set{R}(\KX_{1}, \ldots, \KX_{k}; \tilde{\D}_{1}, \ldots, \tilde{D}_{k})
\end{equation*}
and consequently, by \eqref{eq:MAC-capacity},
\begin{equation*}
  \set{C}(\Es_1, \ldots, \Es_k; \D_1, \ldots, \D_k)
  \supseteq 
  \set{C}(\Es_1, \ldots, \Es_k; \tilde{\D}_1, \ldots, \tilde{\D}_k).
\end{equation*}

We can strengthen this result using the symmetry of the problem as in
the single-user case.  The utility of the rotational
invariance~\eqref{eq:rotI} is, however, rather limited since the LOS
matrices cannot be assumed to be jointly diagonalizable.  Thus, the
monotonicity cannot be simply stated in terms of the singular values
of LOS matrices.  Instead, we have the following.
\begin{corollary}
  \label{cor:mac_contained}
  Let $\D = [\D_1, \ldots, \D_k]$ and $\tilde{\D} = [\tilde{\D}_1,
  \ldots, \tilde{\D}_k]$ be LOS matrices such that
  \begin{equation*}
    \hermi{[{\D}_1 \mat{U}_1, \ldots, {\D}_k \mat{U}_k]}
    [{\D}_1 \mat{U}_1, \ldots, {\D}_k \mat{U}_k]
    \geql \hermi{\tilde{\D}} \tilde{\D}
  \end{equation*}
  for some $\mat{U}_i \in \Unitary{n_i}, i = 1,\ldots, k.$ Then
  \begin{equation*}
    \set{C}(\Es_1, \ldots, \Es_k; \D_1, \ldots, \D_k)
    \supseteq 
    \set{C}(\Es_1, \ldots, \Es_k; \tilde{\D}_1, \ldots, \tilde{\D}_k).
  \end{equation*}
\end{corollary}

%%%%%%%%%%%%%%%%%%%%%%%%%%%%%%%%%%%%%%%%%%%%%%%%%%%%%%%%%%%%%%%%%%%%%%%%

\section{Proof of Theorem~\ref{th:main}}
\label{sec:proof-monotonicity}

Recall that given any $\KX \in \Positive{n}$ and $\D, \tilde{\D} \in
\Complex^{m\times n}$ satisfying
\begin{equation}
  \label{orig-cond}
  \hermi{{\D}} {\D} \geql \hermi{\tilde{\D}} \tilde{\D}
\end{equation}
we wish to show that for all $t \ge 0$,
\begin{equation}
  F(t ; \KX, {\D}) \leq F(t; \KX, \tilde{\D})
  \label{ineq-thm}
\end{equation}
where
\begin{equation*}
  F(t; \KX, \D) = \Prv{ \log\det\left( \mat{I}_m +
      (\Hmat + \D) \KX \hermi{(\Hmat + \D) } \right) \leq t }.
\end{equation*}

Without loss of generality, we can assume that the matrices
$\D$ and $\tilde{\D}$ satisfy
\begin{equation}
  \tilde{\D} = \mat{\Phi}{\D}, \quad
  \mat{\Phi} = \diag{\alpha, 1, \ldots, 1}
  \label{simple-cond}
\end{equation}
for some $0 \le \alpha \le 1$.  We justify this reduction as follows.
Suppose that the desired inequality~\eqref{ineq-thm} holds under the
condition~\eqref{simple-cond}.  Then from the rotational
invariance~\eqref{eq:rotF}, for any permutation matrix $\mat{P}$,
\begin{IEEEeqnarray}{rCl}
  \nonumber
  F(t; \KX, \mat{P}\mat{\Phi}\hermi{\mat{P}}{\D})
  &=& F(t; \KX, \mat{\Phi} \hermi{\mat{P}} {\D}) \\
  \nonumber
  &\ge& F(t; \KX, \hermi{\mat{P}} {\D}) \\
  &=& F(t; \KX, {\D})
  \label{eq:one-dim-scaling}
\end{IEEEeqnarray}
and consequently the result must also hold when $\mat{\Phi} =
\diag{1, \ldots, 1, \alpha, 1, \ldots, 1}$. Expressing
$\diag{\alpha_1, \ldots, \alpha_m}$ as a product
\begin{equation*}
  \diag{\alpha_1, \ldots, \alpha_m} = \diag{\alpha_1, 1, \ldots, 1}
  \cdot \diag{1, \alpha_2, 1, \ldots, 1} \cdot \ldots \cdot
  \diag{1, \ldots, 1, \alpha_m} 
\end{equation*}
and applying the inequality \eqref{eq:one-dim-scaling} $m-1$ times
yields that the result \eqref{ineq-thm} must also hold for any $\D$
and $\tilde{\D}$ such that $\tilde{\D} = \mat{\Phi} {\D}$ with
arbitrary diagonal contraction matrix $\mat{\Phi}$ with $0 \le
\mat{\Phi}_{ii} \le 1,$ $i = 1,\ldots, m$.  Now applying the
rotational invariance~\eqref{eq:rotF} once again to arbitrary unitary
matrices $\mat{U} \in \Unitary{m}$, $\mat{V} \in \Unitary{m}$ and
nonnegative diagonal contraction matrix $\mat{\Phi}$, we obtain
\begin{IEEEeqnarray*}{rCl}
  F(t; \KX, \mat{U}\mat{\Phi}\hermi{\mat{V}}\D)
  &=& F(t; \KX, \mat{\Phi} \hermi{\mat{V}} \D) \\
  &\ge& F(t; \KX, \hermi{\mat{V}} \D) \\
  &=& F(t; \KX, \D).
\end{IEEEeqnarray*}
Thus the desired inequality~\eqref{ineq-thm} holds for any
$\D,\tilde{\D},$ and $\mat{\Phi}$ such that 
\begin{equation}
  \tilde{\D} = \mat{\Phi} {\D},\quad
  \hermi{\mat{\Phi}} \mat{\Phi} \leql \I_m.
  \label{eq:contraction}
\end{equation}
But \eqref{eq:contraction} is equivalent to the original
condition~\eqref{orig-cond} (see, for example, \cite{eaton84}).
Therefore, in order to prove the theorem, it suffices to establish the
inequality~\eqref{ineq-thm} under the simplified
condition~\eqref{simple-cond}.

For the rest of our discussion, we need the following result by
T.~W.~Anderson~\cite{anderson55}~\cite[Theorem 8.10.5]{anderson03}. 
\begin{lemma}
  \label{lem:anderson}
  (Anderson's Theorem) Let $\set{H}$ be a convex set in $\Complex^n$,
  symmetric about the origin (i.e., $\vect{\xi} \in \set{H}$
  implies $-\vect{\xi} \in \set{H}$).  Let $f(\vect{\xi})
  \ge 0$ be a function on $\Complex^n$ such that
  \emph{(i)}
  $f(-\vect{\xi}) = f(\vect{\xi})$ for all $\vect{\xi}$,
  \emph{(ii)} 
  the set~$\{\vect{\xi} \in \Complex^n:\; f(\vect{\xi})
  \geq u\}$ is convex for every $u > 0$; and 
  \emph{(iii)}
  $\int_{\set{H}}
  f(\vect{\xi}) \d \vect{\xi} < \infty$.
  Then
  \begin{equation}
    \int_{\set{H}} f(\vect{\xi}+\alpha\vect{\eta}) \d \vect{\xi}
    \geq \int_{\set{H}} f(\vect{\xi}+\vect{\eta}) \d
    \vect{\xi} 
  \end{equation}
  for every vector~$\vect{\eta} \in \Complex^n$ and $0 \leq \alpha
  \leq 1$.
\end{lemma}

The proof of this celebrated result is based on the Brunn-Minkowski
inequality~\cite{gardner02}.  An interested reader can refer to a nice
review by Perlman~\cite{perlman90} for further generalizations and
applications in multivariate statistics.

Returning to our problem, for any $t \ge 0$, we define a set of
matrices
\begin{equation}
  \set{G}_{t} = \left\{ \mat{G} \in \Complex^{m \times n}:\; \log \det
    \left( \I_m + \mat{G}\KX\hermi{\mat{G}} \right)
    \leq t \right\}.
\end{equation}
For any fixed vectors~$\vect{g}_2, \ldots, \vect{g}_m \in
\Complex^{n}$, let 
\begin{equation}
  \set{H}_{t}\left(\vect{g}_2,\ldots,\vect{g}_m\right) = 
  \left\{ \vect{\xi} \in \Complex^n :\; 
    \trans{[\vect{\xi}, \vect{g}_2, \ldots, \vect{g}_m]}
    \in \set{G}_{t} \right\}.
\end{equation}
In other words, $\set{H}_{t}\left(\vect{g}_2,\ldots,\vect{g}_m\right)$
is the set of the first rows $\trans{\vect{\xi}}$ that belong to
$\set{G}_t$ with given values of other rows
$\trans{\vect{g}_2},\ldots, \trans{\vect{g}_m}$.  As will be checked
later at the end of this section, for any
$\vect{g}_2,\ldots,\vect{g}_m$, the set
$\set{H}_{t}\left(\vect{g}_2,\ldots,\vect{g}_m\right)$ is convex and
symmetric about the origin.

The rest of the proof proceeds along the lines similar to those of Das
Gupta, Anderson, and Mudholkar~\cite{dasguptaandersonmudholkar64}.  We
represent $\Hmat$ as $\trans{[\vect{H}_1, \ldots, \vect{H}_{m}]}$,
where $\trans{\vect{H}}_j$ is the $j$-th row of $\Hmat$.  Similarly,
let $\trans{\vect{d}_{j}}$ denote the $j$-th row of $\D$.  Let
$f(\vect{\xi}| \vect{h}_2,\ldots,\vect{h}_m)$ be the conditional
density of $\vect{H}_{1}$ conditioned on $\vect{H}_j = \vect{h}_j,
j=2,\ldots,m.$ Since the rows of $\Hmat$ are mutually independent,
$f(\vect{\xi}| \vect{h}_2,\ldots,\vect{h}_m) = f(\vect{\xi})$ is
multivariate Gaussian $\NormalC{\vect{0}}{\mat{I}_n}$, which satisfies
the conditions (i) to (iii) of Anderson's Theorem.  Combining the
conditions on $f$ and $\set{H}_t$ with the standing
assumption~\eqref{simple-cond}, we can invoke Anderson's Theorem for
the first row of $\Hmat$ after conditioning on the other rows
$\trans{\vect{H}_2},\ldots, \trans{\vect{H}_m}$ as follows:
\begin{IEEEeqnarray}{rCl}
  \IEEEeqnarraymulticol{3}{l}{ \Prv{\log\det\left(\I_m +
        \left( \Hmat + \D \right) \KX \hermi{\left(
        \Hmat + \D \right)}\right) \leq t \; \bigg| \; 
      \vect{H}_{i} = \vect{h}_{i}, \; i = 2,\ldots, m}}
  \nonumber \\
  &=& \int_{\set{H}_{t}(\vect{h}_{2} + \vect{d}_2, \ldots,                                           
  \vect{h}_m + \vect{d}_m)} f(\vect{\xi} -
  \vect{d}_1) \d \vect{\xi} \nonumber \\
  &\leq& \int_{\set{H}_{t}(\vect{h}_{2} + \vect{d}_2, \ldots,
  \vect{h}_m + \vect{d}_m)} f(\vect{\xi} - \alpha
  \vect{d}_1) \d \vect{\xi} \nonumber \\
  &=& \Prv{\log\det\left(\I_m + \left( \Hmat +
      \tilde{\D} \right) \KX \hermi{\left( \Hmat +
      \tilde{\D} \right)}\right) \leq t \; \bigg| \; 
      \vect{H}_{i} = \vect{h}_{i}, \; i = 2,\ldots, m}. 
\label{eq:cond-prob}
\IEEEeqnarraynumspace
\end{IEEEeqnarray}
By taking the expectation on both sides of \eqref{eq:cond-prob}
with respect to the joint density of $\vect{H}_2, \ldots, \vect{H}_m$,
we establish the desired inequality~\eqref{ineq-thm}.

It remains to check the convexity and symmetry of the set $\set{H}_t =
\set{H}_t(\vect{g}_2,\ldots,\vect{g}_m)$.  Let $\mat{G} =
\trans{[\vect{\xi}, \vect{g}_2,\ldots,\vect{g}_m]}$.  We show that
$\det(\mat{I}_m + \mat{G}\KX\hermi{\mat{G}})$ is convex and
symmetric in $\vect{\xi}$, which clearly implies the convexity and 
symmetry of $\set{H}_t$.  For the symmetry, observe that 
\begin{equation*}
\det(\mat{I}_m + \mat{G}\KX\hermi{\mat{G}})
= \det(\mat{I}_m + \mat{U}\mat{G}\KX\hermi{\mat{G}}\hermi{\mat{U}})
\end{equation*}
for any unitary matrix $\mat{U}$; in particular, $\mat{U}=
\diag{-1,1,\ldots,1}.$

For the convexity, let $\mat{F} = \mat{G}\KX^{\frac{1}{2}}$ where
$\KX^{\frac{1}{2}}$ is any matrix satisfying $\KX^{\frac{1}{2}}
\hermi{(\KX^{\frac{1}{2}})} = \KX$.  Recall the identity
\begin{equation}
  \label{eq:detAB}
  \det{(\I_k + \mat{A}\mat{B}}) = \det{(\I_j + \mat{B}\mat{A})}
\end{equation}
for any $\mat{A} \in \Complex^{k \times j}, \mat{B} \in \Complex^{j
\times k}$.  Then we have
\begin{IEEEeqnarray}{rCl}
  \nonumber \det\left(\I_m + \mat{G}\KX\hermi{\mat{G}} \right) &=&
  \nonumber \det\left(\I_m + \mat{F}\hermi{\mat{F}} \right) \\
  &=&
  \nonumber \det\left(\I_n + \hermi{\mat{F}}\mat{F} \right) \\
  &=& 
  \nonumber
  \det\left( \I_n + \sum_{j=2}^m \hermi{(\trans{\vect{f}_j})}
    \trans{\vect{f}_j} +
    \hermi{(\trans{\vect{f}_1})} \trans{\vect{f}_1} \right) \\
  \nonumber    
  &=& \det\left( \mat{M} + \hermi{(\trans{\vect{f}_1})}
    \trans{\vect{f}_1} \right) \\
  \nonumber
  &=& \det\left( \mat{M} \right) \det\left( \I_n + \mat{M}^{-1}
    \hermi{(\trans{\vect{f}_1})} \trans{\vect{f}_1} \right) \\
  \nonumber
  &=& \det\left( \mat{M} \right) \left( 1 + \trans{\vect{f}_1}
    \mat{M}^{-1} \hermi{(\trans{\vect{f}_1})} \right) \\
  &=& \det\left( \mat{M} \right) \left( 1 + \trans{\vect{\xi}}\KX^{\frac{1}{2}}
    \mat{M}^{-1} \KX^{\frac{1}{2}}\hermi{(\trans{\vect{\xi}})} \right)
  \label{eq:quad}
\end{IEEEeqnarray}
where $\trans{\vect{f}_j}$ denotes the $j$-th row of $\mat{F}$ and the
positive definite matrix $\mat{M}$ is defined as $\mat{M} = \I_n +
\sum_{j=2}^m \hermi{(\trans{\vect{f}_j})} \trans{\vect{f}_j}$.  The
last line of \eqref{eq:quad} is a positive semidefinite quadratic form
in $\vect{\xi}$, and hence it is convex.

\section{Concluding Remarks}

In this paper we have found a natural ordering of MIMO Rician channels
via their LOS matrices. We have shown that for two LOS matrices $\D,
\tilde{\D} \in \Complex^{m \times n}$
\begin{equation}
\nonumber
  \hermi{\D} \D \geql \hermi{\tilde{\D}} \tilde{\D}
  \Longleftrightarrow
  \Big( \IF(\KX, \D) \geq \IF(\KX, \tilde{\D}) \quad
  \forall \KX \in \Positive{n} \Big)
\end{equation}
where $\IF(\KX, \D) = I(\X; \Y | \Hmat)$ is the mutual information
induced by a $\NormalC{\vect{0}}{\KX}$ input over a coherent MIMO
Rician channel with LOS matrix $\D$. From this result we obtained
monotonicity results for isotropic Gaussian input information rate and
for channel capacity, not only for the single-user channel but also
for the multiple-access channel.

In some sense the results of this paper may not be surprising because
the relation $\hermi{\D} \D \geql \hermi{\tilde{\D}} \tilde{\D}$
implies $\trace{\hermi{\D} \D} \geq \trace{\hermi{\tilde{\D}}
  \tilde{\D}}$ and hence a larger output power. Note, however, that
some care must be exercised because in MIMO communications a larger
output power need not imply a larger capacity. For instance, if
\begin{IEEEeqnarray*}{rClrCl}
  \D_1 &=&
  \begin{pmatrix}
    10 & 10 \\ 10 & 0
  \end{pmatrix},\quad
  &
  \D_2 &=&
  \begin{pmatrix}
    10 & 10 \\ 10 & 10
  \end{pmatrix}
\end{IEEEeqnarray*}
then although the power in the LOS component increases while changing
from $\D_1$ to $\D_2$, one can numerically show that the isotropic
Gaussian input information rate and channel capacity are larger on the
channel with LOS matrix $\D_1$ than on the channel with LOS matrix
$\D_2$. The intuition is that $\D_1$ has full rank with singular
values $16.18$ and $6.18$, whereas $\D_2$ is rank deficient with
singular values $20$ and $0$, thus providing only one LOS eigenmode.

%%%%%%%%%%%%%%%%%%%%%%%%%%%%%%%%%%%%%%%%%%%%%%%%%%%%%%%%%%%%%%%%%%%

\appendix

\section{Proof of Proposition~\ref{prop:converse}}
\label{sec:proof-prop-refpr}

Instead of proving Proposition~\ref{prop:converse} directly, we will
prove the equivalent statement
\begin{equation*}
  \hermi{{\D}} {\D} \nsucceq \hermi{\tilde{\D}} \tilde{\D} \quad
  \Rightarrow \quad
  \IF(\KX, \D) < \IF(\KX, \tilde{\D}) \ \text{for some}\ \KX \in \Positive{n}.
\end{equation*}
We first note that $\hermi{{\D}} {\D} \nsucceq \hermi{\tilde{\D}}
\tilde{\D}$ means that there exists a vector $\vect{a} \in \Complex^n$
such that
\begin{equation}
  \label{eq:noncentrality}
  \hermi{\vect{a}} \hermi{{\D}} {\D} \vect{a} <
  \hermi{\vect{a}} \hermi{{\tilde{\D}}} \tilde{\D} \vect{a}.  
\end{equation}
For such a vector $\vect{a}$, let $\KX_0 = \vect{a} \hermi{\vect{a}}
\in \Positive{n}$. We will show that for $\KX_0$ the strict inequality
$\IF(\KX_0, \D) < \IF(\KX_0, \tilde{\D})$ holds.

By \eqref{eq:I_to_F} it suffices to show that $F(t; \KX_0, \D) > F(t;
\KX_0, \tilde{\D})$ for all $t > 0$.  Define
$\vect{G} = \Hmat \vect{a}$, $\vect{b} = \D \vect{a}$, and
$\tilde{\vect{b}} = \tilde{\D}\vect{a}$.  Then we have  
for any $t > 0$
\begin{IEEEeqnarray}{rCl}
  \nonumber
  F(t; \KX_0, \D) &=& \Prv{ \log\det\left( \mat{I}_m + (\Hmat + \D)
      \vect{a} \hermi{\vect{a}} \hermi{(\Hmat + \D) } \right) \leq t }
  \\
  \label{eq:detAB2}
  &=& \Prv{ \log\left( 1 + \hermi{\vect{a}} \hermi{(\Hmat + \D) }
      (\Hmat + \D) \vect{a} \right) \leq t } \\
  \nonumber
  &=& \Prv{ \log\left( 1 + \hermi{(\vect{G} + \vect{b})} (\vect{G} +
      \vect{b}) \right) \leq t } \\
  \label{eq:monotonicitynoncentralchisquare}
  &>& \Prv{ \log\left( 1 + \hermi{(\vect{G} + \tilde{\vect{b}})}
      (\vect{G} + \tilde{\vect{b}}) \right) \leq t } \\
  \nonumber
  &=& \Prv{ \log\left( 1 + \hermi{\vect{a}} \hermi{(\Hmat +
        \tilde{\D}) } (\Hmat + \tilde{\D}) \vect{a} \right) \leq t }
  \\
  \nonumber
  &=& \Prv{ \log\det\left( \mat{I}_m + (\Hmat + \tilde{\D}) \vect{a}
      \hermi{\vect{a}} \hermi{(\Hmat + \tilde{\D}) } \right) \leq t }
  \\
  \nonumber
  &=& F(t; \KX_0, \tilde{\D})
\end{IEEEeqnarray}
where \eqref{eq:detAB2} follows from \eqref{eq:detAB} and
\eqref{eq:monotonicitynoncentralchisquare} follows from the
strict monotonicity result for the single-antenna
case~\cite[Lemma~6.2~(b)]{lapidothmoser03_3}.  Indeed, $\vect{G}$ is
distributed according to
$\NormalC{\vect{0}}{\hermi{\vect{a}}\vect{a}\, \I_m}$ and
$\hermi{(\vect{G} + \vect{b})} (\vect{G} + \vect{b})$ has a scaled
non-central chi-square distribution with (scaled) non-centrality
parameter $\hermi{\vect{b}}\vect{b}$.  Now $\hermi{(\vect{G} +
\tilde{\vect{b}})} (\vect{G} + \tilde{\vect{b}})$ in
\eqref{eq:monotonicitynoncentralchisquare} is also a scaled
non-central chi-square random variable, which, from
\eqref{eq:noncentrality}, has a strictly larger non-centrality
parameter $\hermi{\tilde{\vect{b}}} \tilde{\vect{b}} >
\hermi{\vect{b}}\vect{b}$. Hence, $\hermi{(\vect{G} +
\tilde{\vect{b}})} (\vect{G} + \tilde{\vect{b}})$ is stochastically
strictly larger than $\hermi{(\vect{G} + \vect{b})} (\vect{G} +
\vect{b})$, so that the strict inequality in
\eqref{eq:monotonicitynoncentralchisquare} is justified for any $t >
0$.

%%%%%%%%%%%%%%%%%%%%%%%%%%%%%%%%%%%%%%%%%%%%%%%%%%%%%%%%%%%%%%%%%%%


\begin{thebibliography}{10}
\providecommand{\url}[1]{#1}
\csname url@rmstyle\endcsname
\providecommand{\newblock}{\relax}
\providecommand{\bibinfo}[2]{#2}
\providecommand\BIBentrySTDinterwordspacing{\spaceskip=0pt\relax}
\providecommand\BIBentryALTinterwordstretchfactor{4}
\providecommand\BIBentryALTinterwordspacing{\spaceskip=\fontdimen2\font plus
\BIBentryALTinterwordstretchfactor\fontdimen3\font minus
  \fontdimen4\font\relax}
\providecommand\BIBforeignlanguage[2]{{%
\expandafter\ifx\csname l@#1\endcsname\relax
\typeout{** WARNING: IEEEtran.bst: No hyphenation pattern has been}%
\typeout{** loaded for the language `#1'. Using the pattern for}%
\typeout{** the default language instead.}%
\else
\language=\csname l@#1\endcsname
\fi
#2}}

\bibitem{lapidothmoser03_3}
A.~Lapidoth and S.~M. Moser, ``Capacity bounds via duality with applications to
  multiple-antenna systems on flat fading channels,'' \emph{IEEE Transactions
  on Information Theory}, vol.~49, no.~10, pp. 2426--2467, October 2003.

\bibitem{johnsonkotzbalakrishnan95_2}
N.~L. Johnson, S.~Kotz, and N.~Balakrishnan, \emph{Continuous Univariate
  Distributions}, 2nd~ed.\hskip 1em plus 0.5em minus 0.4em\relax John Wiley \&
  Sons, 1995, vol.~2.

\bibitem{cottatelluccidebbah04}
L.~Cottatellucci and M.~Debbah, ``The effect of line of sight on the asymptotic
  capacity of {MIMO} systems,'' in \emph{Proceedings IEEE International
  Symposium on Information Theory (ISIT)}, Chicago, Illinois, USA, June 27 --
  July 2, 2004, p. 241.

\bibitem{jayaweerapoor05}
S.~K. Jayaweera and H.~V. Poor, ``On the capacity of multiple-antenna systems
  in {R}ician fading,'' \emph{IEEE Transactions on Wireless Communications},
  vol.~4, no.~3, pp. 1102--1111, May 2005.

\bibitem{lebrunfaulknershafismith04}
G.~Lebrun, M.~Faulkner, M.~Shafi, and P.~J. Smith, ``{MIMO} {R}icean channel
  capacity,'' \emph{2004 IEEE International Conference on Communications},
  vol.~5, pp. 2939--2943, 2004.

\bibitem{driessenfoschini99}
P.~Driessen and G.~Foschini, ``On the capacity formula for multiple
  input-multiple output wireless channels: A geometric interpretation,''
  \emph{IEEE Transactions on Communications}, vol.~47, no.~2, pp. 173--176,
  February 1999.

\bibitem{hoeslilapidoth04}
D.~H\"osli and A.~Lapidoth, ``The capacity of a {MIMO} {Ricean} channel is
  monotonic in the singular values of the mean,'' in \emph{Proceedings of the
  5th International ITG Conference on Source and Channel Coding (SCC)},
  Erlangen, Germany, January 14--16, 2004, pp. 381--385.

\bibitem{venkatesansimonvalenzuela03}
S.~Venkatesan, S.~H. Simon, and R.~A. Valenzuela, ``Capacity of a {G}aussian
  {MIMO} channel with nonzero mean,'' in \emph{Proceedings of the IEEE
  Semiannual Vehicular Technology Conference}, Orlando, FL, October 6--9 2003,
  pp. 1767--1771.

\bibitem{wyner94}
A.~D. Wyner, ``{S}hannon-theoretic approach to a {G}aussian cellular
  multiple-access channel,'' \emph{IEEE Transactions on Information Theory},
  vol.~40, no.~6, pp. 1713--1727, November 1994.

\bibitem{shamaiwyner97}
S.~Shamai~(Shitz) and A.~D. Wyner, ``Information-theoretic considerations for
  symmetric, cellular, multiple-access fading channels --- part~{I},''
  \emph{IEEE Transactions on Information Theory}, vol.~43, no.~6, pp.
  1877--1894, November 1997.

\bibitem{eaton84}
M.~L. Eaton, ``On group induced orderings, monotone functions, and convolution
  theorems,'' in \emph{Inequalities in Statistics and Probability}, ser.
  Lecture Notes --- Monograph Series, Y.~L. Tong, Ed., vol.~5.\hskip 1em plus
  0.5em minus 0.4em\relax Institute of Mathematical Statistics, Hayward,
  California, 1984, pp. 13--25.

\bibitem{anderson55}
T.~W. Anderson, ``The integral of a symmetric unimodal function over a
  symmetric convex set and some probability inequalities,'' \emph{Proceedings
  of the American Mathematical Society}, vol.~6, no.~2, pp. 170--176, 1955.

\bibitem{anderson03}
------, \emph{An Introduction to Multivariate Statistical Analysis},
  3rd~ed.\hskip 1em plus 0.5em minus 0.4em\relax John Wiley \& Sons, 2003.

\bibitem{gardner02}
R.~Gardner, ``The {B}runn-{M}inkowski inequality,'' \emph{Bulletin of the
  American Mathematical Society}, vol.~39, no.~3, pp. 355--405, 2002.

\bibitem{perlman90}
M.~D. Perlman, ``{T.~W.~Anderson's} theorem on the integral of a symmetric
  unimodal function over a symmetric convex set and its applications in
  probability and statistics,'' in \emph{The Collected Papers of
  T.~W.~Anderson}, G.~P.~H. Styan, Ed.\hskip 1em plus 0.5em minus 0.4em\relax
  Wiley, 1990, vol.~2.

\bibitem{dasguptaandersonmudholkar64}
S.~Das~Gupta, T.~W. Anderson, and G.~S. Mudholkar, ``Monotonicity of the power
  functions of some tests of the multivariate linear hypothesis,'' \emph{The
  Annals of Mathematical Statistics}, vol.~35, no.~1, pp. 200--205, March 1964.

\end{thebibliography}
\end{document}